\documentclass[letterpaper,journal]{IEEEtran}

\def \sn {\textnormal{\textsc{APEX}}}
\usepackage{algorithm}
\usepackage{algorithmicx}  \usepackage{algpseudocode} \usepackage{amsfonts}
\usepackage{amsmath}
\usepackage{textcomp}
\usepackage{array}
\usepackage{xcolor}
\usepackage{epsfig}
\PassOptionsToPackage{hyphens}{url}
\usepackage{caption}
\usepackage{color}
\usepackage{tcolorbox} \usepackage{listings}
\tcbuselibrary{skins}

\usepackage{dingbat}
\usepackage{stfloats}
\usepackage{tabularx}
\usepackage{booktabs}
\usepackage{mdwmath}
\usepackage{multirow}
\usepackage{enumerate}

\usepackage{amssymb}

\usepackage{graphicx}
\usepackage{wasysym}
\usepackage{subfigure}
\usepackage{mathrsfs}

\usepackage{multirow}
\usepackage{url}

\usepackage[hidelinks]{hyperref}
\hypersetup{pdftitle={Automated Post-Exploitation Attack Chain Exploration Based on a Phase-guided Hybrid Planning-Monte Carlo Tree Search Algorithm},pdfauthor={Jie Ying, Mengce Zheng, Ruoxi Chen, Wenrui Cheng, Tiantian Zhu, Zhongjie Zhu}}

\begin{document}

\title{Automated Post-Exploitation Attack Chain Exploration Based on a Phase-guided Hybrid Planning-Monte Carlo Tree Search Algorithm}

\author{Jie Ying,
        Mengce Zheng,
        Ruoxi Chen,
        Wenrui Cheng,
        Tiantian Zhu,
        Zhongjie Zhu*
		
\IEEEcompsocitemizethanks{
\IEEEcompsocthanksitem This work has been submitted to the IEEE for possible publication. Copyright may be transferred without notice, after which this version may no longer be accessible.
\IEEEcompsocthanksitem This work is supported in part by the following grants: National Natural Science Foundation of China under Grant 62002324. Zhejiang Provincial Natural Science Foundation of China under Grant No. LQ21F020016. The Fundamental Research Funds for the Provincial Universities of Zhejiang under Grant No. RF-A2023009. Key Program of Zhejiang Province Natural Science Foundation (LZ24F010004), Zhejiang Provincial "Pioneer, Leader+ X" Science and Technology Program (2026C02A1221), The 14th China Innovation and Entrepreneurship Competition-Innovation Challenge (Ningbo).
\IEEEcompsocthanksitem J. Ying, M. Zheng, R. Chen, and Z. Zhu* are with the College of Information and Intelligent Engineering, Zhejiang Wanli University, Ningbo 315100, China. E-mail: yingjie@zwu.edu.cn, mczheng@zwu.edu.cn, chenruoxi@zwu.edu.cn, zhongjiezhu@hotmail.com*, *corresponding author. 
\IEEEcompsocthanksitem T. Zhu and W. Cheng are with the College of Computer Science and Technology, Zhejiang University of Technology, Hangzhou 310023, China. E-mail: ttzhu@zjut.edu.cn, zjutcwr@zjut.edu.cn. 
}
}

\markboth{APEX: Automated Post-Exploitation Attack Chain Exploration}{Ying et al.: Automated Post-Exploitation Attack Chain Exploration}

\maketitle

\begin{abstract}
Advanced attackers in post-exploitation scenarios increasingly craft attack chains that combine benign operations, configuration errors, and vulnerability exploitation to achieve objectives like privilege escalation. While such chains pose significant threats, manually discovering them in complex modern host environments is prohibitively labor-intensive. Automating this discovery is therefore essential for proactive defense, yet existing methods still suffer from inefficiency and lack diversity when exploring these attack chains.
\par
To address these limitations, we present \sn{}, an Automated Post-EXploitation attack chain exploration system. This system transforms attack chain discovery into a planning problem and solves it with a novel phase-guided hybrid Planning-Monte Carlo Tree Search (MCTS) algorithm. By organizing attack actions into semantic phases and integrating symbolic planning with MCTS, \sn{} enables efficient exploration of the state space while generating multiple diverse chains. Evaluations on a benchmark comprising 5 virtual machines, 30 synthetic problems and 291 cloud instances demonstrate that \sn{} successfully discovers both known and previously unreported attack chains—including 9 novel privilege escalation vectors—providing actionable intelligence for hardening complex production environments.
\end{abstract}

\begin{IEEEkeywords}
Attack Chain Exploration, Post-Exploitation Attack, AI Planning, Monte Carlo Tree Search Algorithm.
\end{IEEEkeywords}

\section{Introduction}
\label{sec:intro}
Over the past decades, to prevent potential attackers from exploiting software vulnerabilities to compromise systems, security researchers have proposed numerous techniques for automated software vulnerability detection, such as static analysis~\cite{chess2004static, baldoni2018survey}, dynamic analysis~\cite{liang2018fuzzing, fioraldi2020afl++, clause2007dytan, newsome2005dynamic, wright2006sandbox}, and hybrid analysis~\cite{rus2002hybrid, roundy2010hybrid}. However, not all vulnerabilities can be exploited by attackers to gain control of a system. The Forum of Incident Response and Security Teams (FIRST)~\cite{first} proposed the Common Vulnerability Scoring System (CVSS)~\cite{cvss} which evaluates and classifies vulnerabilities. Among these, only vulnerabilities rated as critical (scoring range: 9.0–10.0) can be remotely exploited without user interaction and have widespread impact. According to reports~\cite{vulnerabilityblog}, only 13.7\% of vulnerabilities were marked as critical. In addition, there are many techniques such as Control Flow Integrity (CFI)~\cite{burow2017control}, Address Space Layout Randomization (ASLR)~\cite{shacham2004effectiveness}, and Shadow Stack~\cite{dang2015performance} to restrict attackers from exploiting vulnerabilities.

Consequently, attackers in post-exploitation scenarios—after gaining initial access with limited privileges—increasingly rely on multi-step attack chains that intertwine benign system operations and malicious actions to achieve phased objectives such as privilege escalation and lateral movement. For example, an attacker might chain configuration enumeration, malicious library crafting, environment variable manipulation, and execution of a misconfigured binary to escalate to root privileges.
\par
Post-exploitation attack chains exhibit three distinctive characteristics: multi-phase progression, target-specific precision, and high operational efficiency. By executing these carefully orchestrated chains, attackers can rapidly expand their operational capabilities and achieve specific objectives. Conversely, defenders can leverage intelligence about these chains to rectify system misconfigurations and deploy software patches promptly. Undoubtedly, these realities highlight the critical necessity for implementing automated post-exploitation attack chain discovery at the host level.
\par
Currently, the identification of post-exploitation attack chains remains a manual process. The inherent complexity of modern operating systems and executable environments makes the task of discovering these attack chains both technically challenging and time-consuming~\cite{maeda2021automating, miller2018automated, de2024chainreactor}. Researchers have discovered that the permission configurations, library file loading mechanisms, and process capabilities of a host system can be formally modeled, allowing the task of attack chain discovery to be equivalently transformed into a feasible solution search task within the planning domain. Meanwhile, planning algorithms have demonstrated significant effectiveness in the field of network attack simulation and modeling~\cite{amos2017efficient, hoffmann2015simulated, miller2018automated, speicher2019towards}, leading researchers to explore their application in solving attack chains discovery~\cite{de2024chainreactor, choi2021plan2defend}.
\par
However, integrating host modeling with planning algorithms for post-exploitation attack chain discovery currently faces three significant challenges:
(1) Modeling Imprecision: Existing approaches typically model hosts as single nodes with attributes (e.g., OS, vulnerabilities, users)~\cite{maeda2021automating, li2025her}, but this coarse-grained representation fails to capture an attacker's evolving capabilities during post-exploitation phases accurately.
(2) Computational Inefficiency: The complexity of host environments results in a huge search space after modeling. Given attackers' limited dwell time on compromised systems~\cite{lateral-movement}, conventional planners often cannot generate viable solutions within operationally relevant time constraints.
(3) Solution Inflexibility: Traditional planning solvers employ satisficing strategies that terminate upon finding any feasible solution~\cite{hoffmann2001ff, smith1998conformant, helmert2006fast, bonet2001planning}. Even when configured for optimal planning (minimizing action steps)~\cite{likhachev2003ara}, these systems cannot provide alternative paths when operational constraints (e.g., failed vulnerability exploitation or file access restrictions) render the primary solution unimplementable.
\par
To address these issues and achieve rapid, feasible, and flexible attack chain exploration, we propose \sn{}, an Automated Post-EXploitation attack chain exploration system. In essence, \sn{} formulates attack chain exploration as an AI planning problem: given the current state of a host (e.g, tree-structured file system, user-group relationships, and specific versions of executable binaries) along with a defined goal (e.g., reading a particular file or escalating to a specific user’s privileges), it generates a sequence of actions (i.e., attack chain) to achieve the objective.
\par
Specifically, we first heuristically define Objects (e.g., files, users, groups, and permissions), Facts (e.g., file readability, user-group membership), and Actions (e.g., executing binaries, modifying file permissions) involved in post-exploitation. This process, known as domain modeling in AI planning, utilizes a formal language to precisely describe the static structure and dynamic possibilities of the target, serving as a prerequisite and universal procedure. 
Subsequently, with low privileges (common in post-exploitation phases), \sn{} automatically retrieves relevant information from the target host and generates a formal host characterization file. 
\par
Then, based on observed phased characteristics in post-exploitation behaviors, we propose a phase-guided hybrid planning-MCTS algorithm. Given the characterization file and a specific objective, \sn{} leverages this algorithm to solve for multiple feasible action chains within the solution space. Finally, we employ a Large Language Model (LLM) to translate the formally described attack chains into natural language reports, providing targeted remediation suggestions to assist security analysts in identifying potential attack vectors and reinforcing critical system components. For example, if an attacker escalates privileges by executing a misconfigured binary, defenders can proactively adjust its permissions to mitigate potential exploit risks.
\par
In this paper, we make the following contributions:
\begin{itemize}
    \item We conduct domain modeling for post-exploitation phases using formal language to describe static structures and dynamic capabilities while maintaining minimalism in the domain model to enhance subsequent planning efficiency.
    \item We propose a phase-guided hybrid Planning-Monte Carlo Tree Search (MCTS) algorithm that leverages the phased nature of post-exploitation behaviors, combining the efficiency of AI planning with the flexibility of MCTS to enable efficient and flexible attack chain exploration.
    \item We implement a prototype of \sn{} and evaluate it on a benchmark comprising 5 virtual machines with known privilege escalation vulnerabilities and 30 synthetic problems, followed by large-scale real-world assessments on 291 live cloud instances, uncovering 9 unreported attack chains.
\end{itemize}

\section{Background}
\label{sec:back}

\subsection{Post-Exploitation in Cyber Attacks}
\label{subsec:post-exploitation}
Post-exploitation activities occur after initial system compromise, focusing on maintaining access, escalating privileges, moving laterally, and achieving objectives while evading detection. In sophisticated attacks like APTs, this phase involves orchestrating multi-step chains that blend malicious actions with legitimate system interactions (e.g., abusing sudo, cron, or environment variables). These chains exploit complex host environments (permissions, services, configurations), making automated exploration crucial for defense. Our work focuses on automatically generating such attack chains from a constrained initial foothold. For a detailed analysis of the relationship with the MITRE ATT\&CK framework and specific attack vectors, see Appendix~\ref{app:mitre-attack-details}.

\subsection{AI Planning}
\label{subsec:ai-planning}
We model post-exploitation as a classical AI planning problem: finding a sequence of actions that transforms an initial system state (compromised host) into a goal state (e.g., root access). We use the Planning Domain Definition Language (PDDL) to formally describe the domain (system objects, facts, actions) and problem (concrete host state, attacker goal). A planning solver then searches for valid action sequences (plans). However, the state space for real systems is vast, and planning from scratch for every variation is computationally expensive. The formal definition of the planning problem and details of our PDDL domain specification are provided in Appendix~\ref{app:planning-formalization}.

\subsection{Monte Carlo Tree Search}
\label{subsec:mcts}
To efficiently navigate the large state space, we leverage Monte Carlo Tree Search (MCTS). MCTS builds a search tree through iterative sampling (selection, expansion, simulation, backpropagation), balancing exploration and exploitation. Its anytime nature suits time-constrained scenarios. Our key insight is to hybridize MCTS with planning: we use phase-based heuristics to guide MCTS's action selection and employ a planner to dynamically generate feasible actions within each phase, dramatically reducing enumeration overhead (which dominates runtime in naive approaches, approximately 91\%). This hybrid approach enables the efficient discovery of diverse and concise attack chains.

\section{Motivation \& Threat Model}
\label{sec:motivation_threat_model}

\subsection{Motivation}
\label{subsec:motivation}
Modern post-exploitation attacks increasingly rely on intricately chaining together seemingly benign system operations with pivotal malicious exploits to form complex attack chains. Successful compromise is rarely the result of a single vulnerability; instead, attackers must strategically combine system misconfigurations, software vulnerabilities, and legitimate operations in specific sequences that create paths to critical security objectives (e.g., root access).
One representative example is a chain combining cron misconfigurations, a vulnerable binary, and file manipulation to reach root privileges. A detailed walkthrough of this example is provided in Appendix~\ref{app:motivation-example}.
\par
Manually discovering these multi-vector chains requires analysts to mentally traverse a vast state space, connecting disparate system attributes (scheduled tasks, binary versions, permissions, etc.) through non-obvious causal relationships. The challenge is magnified by the scale of modern environments—hundreds of software packages, numerous services, and complex permission hierarchies—creating a state space with millions of possibilities. Existing automated tools are often ill-suited to reason about these complex, stateful chains that blend legitimate and malicious operations.
\par
The operational necessity for automated attack chain exploration is clear. Advanced Persistent Threats (APT) increasingly employ techniques that combine legitimate system functionalities ("Living-off-the-Land") with precise exploitation. Proactive defense therefore demands tools capable of proactively discovering such multi-step vectors before they are weaponized. For practitioners, this capability enables preventative hardening through systematic patching, configuration changes, and permission adjustments.

\subsection{Threat Model}
\label{subsec:threat_model}
Our threat model centers on the post-exploitation phase, where an attacker has already gained initial shell access (e.g., as a low-privileged user) on a Unix-based system. The core objective of this work is to automatically explore feasible attack chains that allow the attacker to achieve a specified goal from this initial state, such as privilege escalation, file tampering, data exfiltration, or lateral movement.

The attacker can leverage existing system utilities for reconnaissance and is capable of executing actions based on discovered information. While the attacker can introduce external tools if a method is found, they do not initially possess other users' credentials, ruling out straightforward authentication-based escalation.

We focus exclusively on full-fledged Unix systems with init systems (e.g., systemd). This is because init systems provide a complex landscape of persistent services, scheduled tasks, and environment configurations that are essential for constructing realistic, multi-step attack chains. Consequently, containerized environments are excluded from our scope, as their isolated nature and ephemeral lifecycle present fundamentally different security boundaries and attack surfaces, which limits the applicability and relevance of the intricate chains our system aims to discover.

\section{Overview\&Principle}
\label{sec:system_overview}

\subsection{The \sn{} Architecture}
\label{sec4.1_archi}
\begin{figure*}[h!t]
\centering
\scalebox{0.95}{
\includegraphics[width=\linewidth]{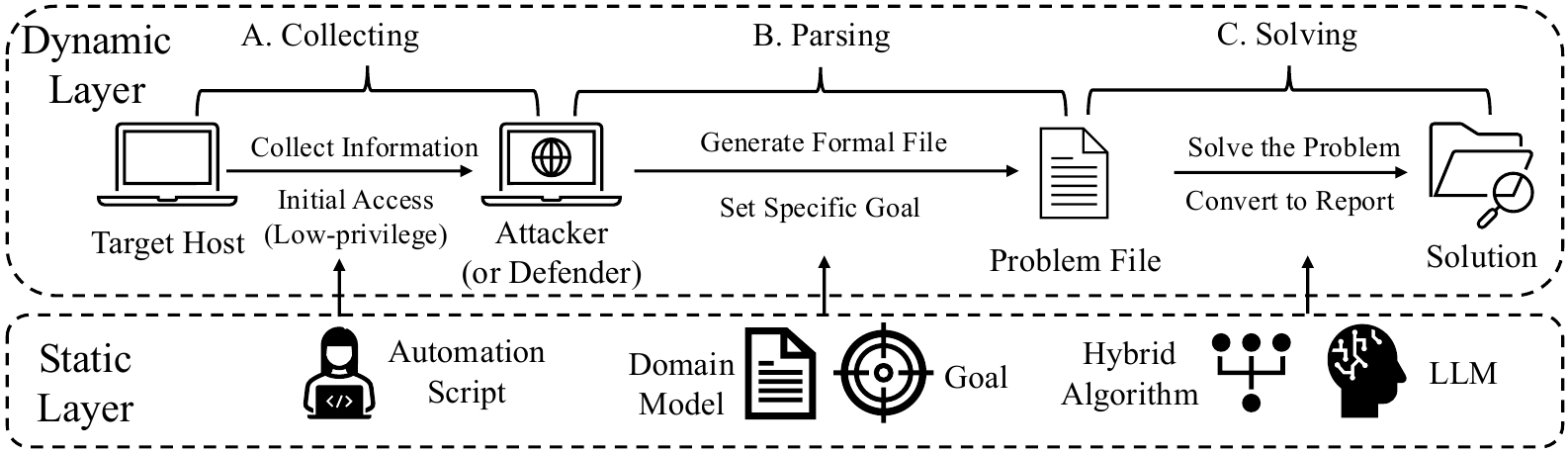}
}
\caption{\sn{} System Architecture.}
\label{fig:system_archi}
\end{figure*}

\sn{} adopts a static-dynamic architecture that systematically integrates static domain knowledge with dynamic host-specific information for automated attack chain discovery in the post-exploitation phase. As illustrated in Figure~\ref{fig:system_archi}, the architecture comprises two complementary layers: a Static Layer and a Dynamic Layer.
\par
The static layer contains the reusable domain model, read-only extraction scripts, and the hybrid Planning-MCTS solver. The dynamic layer applies these components to a target host through three stages: collecting security-relevant host information, parsing the result into a host-specific PDDL problem, and solving for feasible attack chains that are finally translated into analyst-facing reports. This separation keeps domain knowledge reusable while adapting the instantiated problem to each target.

\par
This architecture ensures a clear separation of concerns: the static components encapsulate domain knowledge and solving strategies, while the dynamic layer adapts this framework to a specific Unix-like environment, generating human-readable security reports.

\subsection{Core Design Principles}
\label{sec4.2_design_principle}
\sn{} follows four design principles:
(1) Formal Abstraction: Distilling system complexity into a precise PDDL model ensures logical soundness. (2) Phase-Guided Search: Incorporating post-exploitation phases with specific constraints improves search efficiency. (3) Hybrid Solving: Combining symbolic planning with probabilistic MCTS balances feasibility and diversity. (4) Minimal Interference: Strict read-only operations ensure passive, safe assessment.
Together, these principles guide the modeling, collection, and solving components of \sn{}. Extended explanations of the four principles and their implementation implications are provided in Appendix~\ref{app:principles-details}.

\section{System Design and Modeling}
\label{sec:system_design}
\sn{} follows a model-first design approach: we first define the formal domain model (Section~\ref{subsec:model_philosophy} and Section~\ref{subsec:domain_model_def}), then create scripts to instantiate it (Section~\ref{subsec:automation_script}), and finally develop the hybrid solver algorithm (Section~\ref{subsec:hybrid_algorithm}). This ensures every collected piece of information serves a defined purpose in planning.
\subsection{Modeling Philosophy and Scope}
\label{subsec:model_philosophy}
\par
 Our modeling philosophy is centered on sufficiency and minimalism: the model must capture realistic post-exploitation chains while remaining compact enough for tractable planning. We therefore model a minimal sufficient set of attack surfaces: file-system permissions and ownership, process execution and privilege inheritance, automated execution mechanisms, dependency relationships, software vulnerabilities, user/group relationships, and data movement capabilities.

\par
A naive model that attempts to capture every system detail would lead to a combinatorial explosion in the state space, rendering automated planning infeasible. Conversely, an overly simplistic model would fail to capture the nuanced interactions that constitute real attack chains. Our modeling approach navigates this tension by adhering to a core principle: model only the entities and relationships that are both necessary and sufficient for achieving objectives through the chaining of benign and malicious actions.
\par
We identify and model the following core attack surfaces, arguing that they constitute a minimal sufficient set for post-exploitation reasoning: (1) File system permissions and ownership; (2) Process execution and privilege inheritance; (3) Automated execution mechanisms; (4)  Dependency relationships; (5) Software vulnerabilities; (6) User and group relationships; (7) Data movement capabilities. 
\par
Together, these attack surfaces provide a minimal but sufficient basis for post-exploitation reasoning. Detailed justifications for the selected attack surfaces and the deliberately omitted system aspects are provided in Appendix~\ref{app:modeling-details}.

\subsection{PDDL Domain Definition}
\label{subsec:domain_model_def}
Building upon the aforementioned philosophy, we formally define the domain model (specified in PDDL) as follows.
\par
\textbf{Types (Objects)}. 9 object types representing the core entities in post-exploitation scenarios. The type system follows a structured inheritance hierarchy: \texttt{file} is the base type for all files, with specialized subtypes including \texttt{executable} (for runnable binaries/scripts), \texttt{sharedfile} (for automatically loaded files like libraries), and \texttt{calledexecutable} (for scheduled tasks). Other fundamental types include \texttt{user}, \texttt{group}, \texttt{process}, \texttt{data}, and \texttt{permission}.
\par
\textbf{Predicates (State Properties)}. 30 predicates capture system state, organized into:
\begin{itemize}
    \item Capability Predicates (11 predicates) describe what executables can perform, either through native functionality or known vulnerabilities.
    \item Privilege and Control Predicates (4 predicates) track the attacker's progress and system privilege structure.
    \item File System Predicates (11 predicates) capture static properties and dynamic state of the file system.
    \item Process Execution Predicate (1 predicate) models the process runtime state.
    \item Derived Capability Predicates (3 predicates) compute effective permissions from base predicates. 
\end{itemize}  
\par
The predicate design follows our minimalism principle. This design choice avoids storing a quadratic number of permission facts (users × files) in the initial state, significantly reducing state space complexity.
\par
\textbf{Actions (State Transitions)}. 19 actions model state transitions:
\begin{itemize}
    \item Process Execution Actions (3 actions) establish and leverage execution contexts.
    \item File Manipulation Actions (7 actions) modify file system state.
    \item System Configuration Actions (4 actions) alter access controls and system state.
    \item Derivation and Capability Propagation Actions (5 actions) automatically infer facts when preconditions are met.
\end{itemize}
This compact yet expressive action set enables modeling of complex multi-step attacks while maintaining the minimalism necessary for efficient planning. Each action represents a distinct, non-redundant capability observed in real-world post-exploitation scenarios, with preconditions and effects carefully crafted to capture security-relevant dependencies. Complete predicate definitions and action specifications are provided in Appendix~\ref{app:pddl_details}.

\subsection{Automation Script for Host State Extraction}
\label{subsec:automation_script}
The automation scripts perform passive reconnaissance to instantiate the domain model with concrete host state data, strictly following the Principle of Minimal Interference. These scripts execute only read-only commands to construct a complete snapshot of the system's security-relevant configuration without altering its state.
\par
The extraction is organized into five categories that directly correspond to the attack surfaces defined in our modeling philosophy (Section~\ref{subsec:model_philosophy} and Section~\ref{subsec:domain_model_def}):
\begin{itemize}
    \item Identity and Privilege Context: Enumerates users and groups from system files (\texttt{/etc/passwd}, \texttt{/etc/group}), parses membership relationships, and identifies the attacker's initial foothold context.
    \item File System State: Discovers files in critical directories, extracts security metadata including permissions, ownership, and special flags (SUID/SGID), and classifies files according to our type hierarchy.
    \item System Dynamics: Identifies automated execution mechanisms (cron jobs, systemd services) and maps dependency chains (shared libraries, shell configurations).
    \item Software Vulnerability Context: Queries executable versions and validates them against a CVE knowledge base to identify exploitable vulnerabilities.
    \item Capability Derivation Primitives: Identifies writable locations to support efficient permission computation, avoiding quadratic storage of permission facts.
\end{itemize}
The scripts' design is strictly governed by the domain model, guaranteeing that every piece of collected information serves a precise purpose in the subsequent planning phase.
This tight mapping allows the extracted host state to be translated directly into planning facts with little intermediate processing. Detailed implementation information and exact predicate mappings are provided in Appendix~\ref{app:scripts-details}.

\begin{algorithm}[t]
\caption{Phase-Guided Hybrid Planning-MCTS (PG-HP-MCTS) - Part 1}
\label{alg:PG-HP-MCTS-part1}
\begin{algorithmic}[1]
\small
\Require{Initial state $s_0$, Goal $G$, Base domain $\mathcal{D}_{base}$, Time limit $T$}
\Ensure{Set of attack chains $\Pi$}
\State $\Pi \gets \emptyset$
\State $root \gets \text{createNode}(state=s_0, phase=\phi_{\text{deriv}}, depth=0, action=\emptyset)$
\State \textsc{InitializeCapabilities}($root$) \Comment{Apply derivation actions}
\State $root.phase \gets \phi_{\text{priv}}$ \Comment{Transition to main search}

\For{each iteration within limit $T$}
    \State $node \gets \textsc{Select}($root$)$ \Comment{UCT tree traversal}
    
    \If{$node.depth < d_{max} \land \neg node.isTerminal$}
        \State \textbf{// Phase-Guided Hybrid Expansion}
        \State $\phi \gets node.phase$
        \State $\mathcal{S}_\phi \gets \textsc{GetPhaseActionSchemas}(\phi, node.action)$
        \State $\mathcal{D}_{temp} \gets \textsc{ConstructTempDomain}(\mathcal{D}_{base}, \mathcal{S}_\phi)$
        \State $\mathcal{P}_{temp} \gets \textsc{ConstructTempProb}(node.state, \mathcal{S}_\phi)$
        \State $\mathcal{A} \gets \textsc{ModifiedPlannerQuery}(\mathcal{D}_{temp}, \mathcal{P}_{temp})$ \Comment{Returns grounded actions}
        
        \If{$\mathcal{A} = \emptyset$}
            \State Mark $node$ as terminal
            \State $node' \gets node$
        \Else
            \State $a \gets \textsc{SampleAction}(\mathcal{A})$ 
            \State $s' \gets \textsc{ApplyAction}(node.state, a)$
            \State $\phi' \gets (\phi + 1) \mod 3$ \Comment{Phase transition}
            \State $node' \gets \text{createNode}(state=s', phase=\phi', parent=node, action=a)$
            \State $node.children \gets node.children \cup \{a: node'\}$
        \EndIf
    \Else
        \State $node' \gets node$
    \EndIf
\algstore{apexmain}
\end{algorithmic}
\end{algorithm}

\begin{algorithm}[t]
\caption{Phase-Guided Hybrid Planning-MCTS (PG-HP-MCTS) - Part 2}
\label{alg:PG-HP-MCTS-part2}
\begin{algorithmic}[1]
\small
\algrestore{apexmain}
\setcounter{ALG@line}{0}
\State \textbf{Continuation from Algorithm~\ref{alg:PG-HP-MCTS-part1}}
    \State \textbf{// Phase-Guided Simulation}
    \State $\Delta \gets 0$, $s_{curr} \gets node'.state$, $\phi_{curr} \gets node'.phase$
    \For{$i \gets 1$ to $d_{rollout}$}
        \State $\mathcal{S}_\phi \gets \textsc{GetPhaseActionSchemas}(\phi_{curr}, \emptyset)$
        \State $\mathcal{D}_{temp} \gets \textsc{ConstructTempDomain}(\mathcal{D}_{base}, \mathcal{S}_\phi)$
        \State $\mathcal{P}_{temp} \gets \textsc{ConstructTempProblem}(s_{curr}, \mathcal{S}_\phi)$
        \State $\mathcal{A}_{sim} \gets \textsc{ModifiedPlannerQuery}(\mathcal{D}_{temp}, \mathcal{P}_{temp})$
        
        \If{$\mathcal{A}_{sim} = \emptyset$ \textbf{or} $\textsc{IsTerminal}(s_{curr})$}
            \State \textbf{break}
        \EndIf
        
        \State $a_{sim} \gets \text{RandomChoice}(\mathcal{A}_{sim})$
        \State $s_{curr} \gets \textsc{ApplyAction}(s_{curr}, a_{sim})$
        \State $\phi_{curr} \gets (\phi_{curr} + 1) \mod 3$
        \State $\Delta \gets \Delta + \textsc{StepReward}(s_{curr}, a_{sim})$
    \EndFor
    \State $\Delta \gets \Delta + \textsc{FinalReward}(s_{curr})$
    
    \State \textsc{Backpropagate}($node'$, $\Delta$)
\EndFor
\State \textsc{ExtractPlans}($root$, $\Pi$) \Comment{Collect terminal node paths}
\State \Return $\Pi$

\Function{GetPhaseActionSchemas}{$\phi$, $actionList$}
    \If{$\phi = \phi_{\text{priv}}$}
        \State \Return $\mathcal{S}_\phi$ \Comment{Schema set for privilege manipulation}
    \ElsIf{$\phi = \phi_{\text{dep}}$}
        \State $a_{prev} \gets actionList[-1]$ \Comment{Last action in sequence}
        \State \Return $\textsc{FindDependentSchemas}(a_{prev})$ \Comment{Schemas dependent on $a_{prev}$'s effects}
    \ElsIf{$\phi = \phi_{\text{goal}}$}
        \State \Return $\mathcal{S}_\phi$ \Comment{Schemas that can directly achieve $G$}
    \EndIf
\EndFunction
\end{algorithmic}
\end{algorithm}

\section{The Phase-Guided Hybrid Planning-MCTS Algorithm}
\label{subsec:hybrid_algorithm}
To address the dual challenges of computational intractability and solution inflexibility in automated attack chain exploration, we propose a novel Phase-Guided Hybrid Planning-Monte Carlo Tree Search (PG-HP-MCTS) algorithm. It integrates symbolic AI planning with stochastic search while organizing actions into semantically coherent phases.

\subsection{Phase Guidance: Modeling Attack Action Semantics}
\label{subsubsec:phase_guidance}
The phase guidance mechanism in PG-HP-MCTS organizes attack actions into four semantically distinct categories that reflect their role in the attack progression. This classification emerges from systematic analysis of our domain model and captures the logical dependencies between different types of operations, enabling structured exploration of the attack space.
\par
The \textbf{CAPABILITY\_DERIVATION\_PHASE} ($\phi_{\text{deriv}}$) serves as the initialization stage, encompassing actions that infer implicit system capabilities and establish foundational execution contexts. These operations reveal hidden system properties without directly modifying the state, providing the necessary knowledge for subsequent attack steps. This phase ensures the search begins with a complete understanding of available capabilities.
\par
The \textbf{PRIVILEGE\_MANIPULATION\_PHASE} ($\phi_{\text{priv}}$) involves actions that alter system permissions and security contexts. These operations modify the security landscape to create new attack surfaces or elevate access rights, often serving as prerequisites for more intrusive operations. The actions in this phase focus on changing ownership, permissions, and privilege levels to reshape the system's security boundaries.
\par
The \textbf{EFFECT\_DEPENDENT\_PHASE} ($\phi_{\text{dep}}$) dynamically constructs its action set based on the specific state changes achieved in the preceding phase. For each action $a_{priv}$ executed in $\phi_{\text{priv}}$, the algorithm identifies all actions whose preconditions intersect with $\text{effects}(a_{priv})$. Formally:

\[
\mathcal{A}_{\text{dep}} = \{a \in \mathcal{A} \mid \text{preconditions}(a) \cap \text{effects}(a_{priv}) \neq \emptyset\}
\]

This design ensures that each phase builds directly upon the effects of the previous one, creating a tight causal coupling in the attack progression. For instance, if a privilege manipulation action establishes file writability, this phase will include file writing actions but exclude operations that don't leverage this newly established capability. The phase is not defined by pre-specified action categories but emerges dynamically from the logical dependencies between actions, embodying the principle that effective attack chains exhibit strong stepwise causality.
\par
The \textbf{GOAL\_DIRECTED\_PHASE} ($\phi_{\text{goal}}$) focuses on actions whose effects contain or directly contribute to the goal predicate. Importantly, these are not necessarily terminal actions but often serve as intermediate steps that advance toward the ultimate objective. A goal-directed action might create conditions enabling further privilege manipulation operations in subsequent cycles.
\par
These phases form a logical progression captured by the transition function $next(\phi)$: $\phi_{\text{deriv}} \rightarrow (\phi_{\text{priv}} \rightarrow \phi_{\text{dep}} \rightarrow \phi_{\text{goal}})^{*}, *\ge 1 $ The cyclical nature, implemented by the modulo operation $(\phi + 1) \mod 3$ in Algorithm~\ref{alg:PG-HP-MCTS-part1} (line 20) and Algorithm~\ref{alg:PG-HP-MCTS-part2} (line 14), reflects real attack patterns where goal-directed actions often create new conditions for further system manipulation. For example, a typical privilege escalation chain might progress as: inferring executable capabilities (capability derivation), changing file permissions (privilege manipulation), exploiting the new permissions to write a malicious payload (effect-dependent), and executing it to gain elevated access (goal-directed). The elevated access then enables discovery of new attack surfaces, continuing the cycle toward the ultimate objective.
\par
This phase system implements semantic pruning through several mechanisms. First, it ensures actions are considered in their logical sequence—effect-dependent operations cannot precede the privilege changes that enable them. Second, it restricts the action space at each node to only those operations relevant to the current phase, significantly reducing the branching factor. Third, it enforces causal dependencies between phases, ensuring coherent attack progression.

\subsection{Hybrid Architecture: Synergizing Planning and MCTS}
\label{subsubsec:hybrid_architecture}
The hybrid nature of PG-HP-MCTS addresses complementary weaknesses of pure planning and pure MCTS approaches through carefully designed integration. Our architecture leverages the distinct strengths of each paradigm while mitigating their respective limitations.
\par
AI planning excels at sound and complete action enumeration, rigorously checking preconditions and parameter bindings to guarantee semantic validity. Modern planners (i.e., FF~\cite{hoffmann2001ff}, Fast Downward~\cite{helmert2006fast}, and Powerlifted~\cite{powerlifted}) optimize the computationally expensive process of instantiating action schemas with concrete objects. Conversely, MCTS offers anytime performance crucial for operational time constraints, dynamically balances exploration and exploitation through the UCT formula~\cite{kocsis2006bandit}, and naturally discovers multiple diverse solutions within a single search rather than stopping at the first viable path.
\par
Our hybrid integration, detailed in Algorithms~\ref{alg:PG-HP-MCTS-part1} and \ref{alg:PG-HP-MCTS-part2}, operates through a novel division of labor. Instead of solving complete planning problems, we dynamically generate phase-specific subproblems (lines 10-13 in Algorithm~\ref{alg:PG-HP-MCTS-part1} ). We maintain a base domain $\mathcal{D}_{\text{base}}$ containing the complete set of objects, predicates, and all possible action schemas. At each search node with phase $\phi$, the algorithm invokes $\textsc{GetPhaseActionSchemas}(\phi, \cdot)$ (line 22-31 in Algorithm~\ref{alg:PG-HP-MCTS-part2}) to obtain the phase-specific action schema set $\mathcal{S}_\phi$. This set is then used to construct a temporary domain $\mathcal{D}_{\text{temp}}$ that includes the objects and predicates from $\mathcal{D}_{\text{base}}$, but restricts the action schemas to $\mathcal{S}_\phi$. This temporary domain is paired with a similarly constructed problem file $\mathcal{P}_{\text{temp}}$ encoding the current state, creating a focused planning subproblem that constrains the planner's search to phase-appropriate actions. When a node requires expansion (lines 9-22 of Algorithm~\ref{alg:PG-HP-MCTS-part1}) or simulation (lines 4-16 of Algorithm~\ref{alg:PG-HP-MCTS-part2}), the algorithm identifies action schemas $\mathcal{S}_\phi$ relevant to the node's phase $\phi$, constructs a temporary domain $\mathcal{D}_{temp}$ only containing actions of $\mathcal{S}_\phi$, builds a temporary problem $\mathcal{P}_{temp}$ restriced with current phase, and invokes our modified planner in single-step enumeration mode.
\par
Besides, we also modify the planner to perform only the first layer of state-space expansion. Traditional planners terminate upon finding a complete plan, but our modified version enumerates all grounded actions with satisfied preconditions and returns this set $\mathcal{A}$ without further search (line 13 in Algorithm~\ref{alg:PG-HP-MCTS-part1}). This provides MCTS with a curated, valid action set at minimal computational cost while maintaining semantic correctness.
\par
The MCTS framework then manages the high-level search strategy: selecting which phase progression paths to explore deeper through UCT selection, balancing exploration of different capability derivation strategies, and aggregating rewards from multi-step simulations to evaluate phase sequences. This division creates synergistic efficiency gains through dual pruning mechanisms. Phase guidance provides semantic pruning by restricting actions to phase-appropriate schemas, while the planner provides logical pruning by filtering to only applicable instantiations. Temporary action sets typically comprise only 23\% (on average) of the complete domain action sets, with a 70\% reduction in branch factor compared to unconstrained planning~\ref{sec:eval_bench_envi}).

\subsection{Algorithm Execution and Innovation Impact}
\label{subsubsec:algorithm_execution}
The execution flow of PG-HP-MCTS demonstrates how our innovations interact to address the core challenges of attack chain exploration. Initialization (lines 1-5 of Algorithm~\ref{alg:PG-HP-MCTS-part1}) begins with capability derivation actions in $\phi_{\text{deriv}}$, automatically enriching the state with inferred facts about system capabilities. This preprocessing addresses the issue of insufficient capabilities during the modeling phase before main search commences.
\par
During the expansion phase (lines 9-22 of Algorithm~\ref{alg:PG-HP-MCTS-part1}), the phase attribute $\phi$ determines which action schemas $\mathcal{S}_\phi$ are considered for exploration. When $\phi = \phi_{\text{priv}}$, $\mathcal{S}_\phi$ includes privilege manipulation actions. When $\phi = \phi_{\text{dep}}$, $\mathcal{S}_\phi$ is dynamically constructed by identifying actions whose preconditions intersect with the effects of the preceding privilege manipulation action. When $\phi = \phi_{\text{goal}}$, $\mathcal{S}_\phi$ comprises goal-directed actions. The modified planner then enumerates concrete instantiations, ensuring parameter bindings are valid and all preconditions are satisfied.
\par
The simulation phase (lines 4-16 of Algorithm~\ref{alg:PG-HP-MCTS-part2}) maintains phase guidance even during random rollouts. Each simulation step queries the planner for phase-appropriate actions, ensuring that random exploration respects the semantic constraints of realistic attack progression. The phase transition $(\phi_{curr} + 1) \mod 3$ enforces the cyclical progression pattern throughout simulation, maintaining the integrity of explored chains while allowing for discovery of novel attack vectors within the phase-constrained space. Finally, we provide a detailed description of the pseudocode for the core function $\textsc{GetPhaseActionSchemas}(\phi, \cdot)$ (line 22-31 in Algorithm~\ref{alg:PG-HP-MCTS-part2}).
\par
This design fundamentally transforms the search problem from an unstructured exploration of $\mathcal{O}(|A|^d)$ possible sequences—where $|A|$ is the total action count and $d$ is depth—to a structured exploration of $\mathcal{O}(\prod_{i=1}^{d}|A_{\phi_i}|)$ sequences, where $|A_{\phi_i}| \ll |A|$ is the phase-restricted action set at depth $i$. With our phase system achieving an average reduction of 74\% in action choice per phase, this represents a search space reduction of approximately $\left(\frac{1}{4}\right)^d$, making deep exploration computationally feasible where brute-force approaches would be intractable.
\par
The PG-HP-MCTS algorithm makes three key contributions to automated attack chain exploration. First, it introduces a semantically-grounded phase system that organizes attack actions into capability derivation, privilege manipulation, effect-dependent, and goal-directed phases, providing structural guidance that prunes implausible paths and focuses search on realistic progressions. Second, it develops a tight planner-MCTS integration where a modified planner efficiently enumerates valid actions constrained by phase, while MCTS manages strategic exploration—combining the completeness of planning with the flexibility of stochastic search. Third, it implements a dual efficiency mechanism where phase guidance reduces the action space semantically while planner-assisted filtering reduces it logically, achieving multiplicative reductions in search space complexity.
\par
This integrated approach directly addresses the second and third challenges identified in Section~\ref{sec:intro}: computational inefficiency through dual pruning and hybrid division of labor, solution inflexibility through MCTS's multi-path exploration and anytime performance.

\subsection{Post-Processing and Report Generation}
\label{subsec:postproc_report}
We filter derivation actions from discovered chains and use an LLM to generate natural language reports with step-by-step walkthroughs and mitigation recommendations. This bridges formal planning output and analyst-facing security guidance. The LLM prompt template and a generated report example are provided in Appendix~\ref{app:report-details}.

\section{Evaluation}
\label{sec:evaluation} 
To assess the efficacy, efficiency, and practicality of \sn{}, we conduct a comprehensive evaluation across multiple dimensions. Our experiments are designed to answer the following key research questions (RQs):
\begin{itemize}
    \item \textbf{RQ1 (Effectiveness)}: Can \sn{} successfully discover known and unknown post-exploitation attack chains with high precision?
    \item \textbf{RQ2 (Efficiency)}: Does the proposed Phase-Guided Hybrid Planning-MCTS (PG-HP-MCTS) algorithm outperform baseline planner and pure MCTS in terms of planning time and state-space exploration?
    \item \textbf{RQ3 (Solution Quality \& Diversity)}: Can \sn{} generate multiple, concise, and semantically diverse attack chains, addressing the inflexibility of traditional planners?
    \item \textbf{RQ4 (Practicality)}: Can \sn{} identify novel, previously unreported attack chains in large-scale, real-world cloud environments?
\end{itemize}   

\subsection{Experimental Setup}
\textbf{Implementation \& System Parameters.}
We implemented \sn{} as a Python-based system comprising approximately 8,000 lines of code. The system supports two primary connectivity modes for post-exploitation scenarios: SSH-based collection and reverse-shell listener mode. In both modes, a lightweight agent performs passive reconnaissance through a curated set of read-only commands to extract system state without altering the host.
\par
For capability modeling, \sn{} maps discovered system binaries to formal planning predicates based on the GTFOBins database~\cite{gtfobins}. When an executable matches a GTFOBins entry (e.g., curl with file-download capability), the system automatically instantiates corresponding predicates such as (download\_file /usr/bin/curl) in the initial state. This enables the planner to reason about Living-off-the-Land techniques using real-world tool functionality.
\par
We implemented the PG-HP-MCTS framework from scratch in Python, with careful optimization of state representation and tree node caching. For the integrated planning component, we selected and modified the Powerlifted planner (commit \texttt{41aaddd}) and modified its interface to operate in a "single-step enumeration mode": given a PDDL domain and problem file, it returns all ground actions applicable in the current state (satisfying all preconditions) but does not proceed with full-plan search. This modification is key to the hybrid architecture.
\par
The PG-HP-MCTS parameters were carefully tuned through extensive preliminary experiments on a separate validation set. The MCTS exploration constant was set to \texttt{C=1.414}, with a rollout depth of 15 and a maximum tree depth of 20. The algorithm terminates after reaching either 8,000 iterations or a 30-minute wall-clock time limit. We designed a composite reward function $R(s,a) = 1_{goal}(s')\cdot100+(1-\frac{len(chain)}{20})\cdot 10+\frac{unique\_action(chain)}{total\_chains}\cdot 5 $, where $1_{goal}$indicates goal state achievement. This function strategically balances three objectives: successful goal attainment (primary reward), plan conciseness (shorter chains are preferred), and action schema diversity (encouraging exploration of different attack vectors).
\par
All experiments were conducted on a dedicated server running Ubuntu 22.04 LTS, equipped with an AMD EPYC 7513 32-Core Processor and 256 GB of RAM. Although \sn{}'s planning core is single-threaded, we parallelized the evaluation of independent target instances across CPU cores to maximize throughput. Each planning task was allocated a maximum of 32 GB of RAM and a 30-minute wall-clock time limit.
\par
\textbf{Benchmark and Dataset.}Our evaluation employs a multi-faceted benchmark:
\begin{itemize}
    \item CTF VMs (Known-Chain Validation): We selected 5 well-established Capture-The-Flag virtual machines from VulnHub and HackTheBox that are specifically designed for privilege escalation training:  Escalate My Privileges~\cite{escalate_my_privileges}(C1),  Lin.Security~\cite{lin_security}(C2),  FriendZone~\cite{friend_zone}(C3),  SolidState~\cite{solid_state}(C4), and Jack-of-All-Trades~\cite{jack_of_all_trades}(C5). These VMs contain documented, multi-step privilege escalation chains combining misconfigurations, SUID binaries, and scheduled tasks, providing a ground truth for validation.
    \item Synthetic Problem (Controlled Complexity): 
    To systematically evaluate the scalability of PG-HP-MCTS, we generated 30 synthetic planning problems with controlled complexity levels by incrementally scaling two key parameters: the number of objects (from 500 to 3000 in steps of 500) and the density of predicates (maintaining 3-4 predicates per object to ensure adequate relational structure). Each problem instance was constructed to guarantee at least one solvable privilege escalation path to measure system scalability.
    \item Real-World Cloud Instances (Unreported  Attack Chain): To assess practical impact, we deployed and tested \sn{} on a large-scale dataset of 291 cloud instances: 175 Amazon EC2 instances and 116 Alibaba Cloud ECS instances. We selected a diverse set of official and marketplace Linux images (Ubuntu 18.04/20.04/22.04, CentOS 7/8, Debian 10/11, Amazon Linux 2). All instances were launched with default configurations, and we accessed them via SSH using the default low-privileged user. For fair comparison, we set the goal as obtaining root user privileges. No modifications were made post-creation.
\end{itemize}
\par
\textbf{Baselines for Comparison.} We compare \sn{} against three critical baselines. (1) Pure Optimal Planner (Powerlifted~\cite{powerlifted}): Represents the state-of-the-art in optimal classical planning, configured for satisficing planning (i.e., seeking a feasible solution rather than an optimal one). (2) Pure MCTS: A standard Monte Carlo Tree Search implementation that does not incorporate phase guidance or planner integration. It performs a naive random rollout for simulation. (3) ChainReactor~\cite{de2024chainreactor}: We faithfully reproduced the core planning pipeline of ChainReactor using its reported configuration and ran it on our identical benchmark datasets for a direct, fair comparison.

\subsection{Evaluation on Benchmark Environments}
\label{sec:eval_bench_envi}
\textbf{Known CTF Virtual Machines.} \sn{} successfully discovered valid privilege escalation chains in all 5 CTF VMs, achieving a 100\% success rate. However, due to the inherent complexity of state-space search and the probabilistic nature of the PG-HP-MCTS algorithm, the discovered chains were not always identical to the documented "ground-truth" solutions. Instead, \sn{} frequently found alternative, functionally equivalent paths that leveraged different combinations of system misconfigurations and capabilities, demonstrating the algorithm's ability to explore diverse regions of the solution space. The average time to first solution was 870.6s, reflecting the realistic computational cost of navigating large, complex state spaces inherent to real host environments.
\par
The results, summarized in Table~\ref{tab:ctf_vm}, highlight a key strength of \sn{} in addressing RQ3 (Solution Quality \& Diversity). On average, \sn{} found 2.4 distinct attack chains per VM. Notably, the number of chains discovered exhibited variability across different runs on the same VM, which is characteristic of the stochastic MCTS search process. Critically, all discovered chains were manually validated and confirmed to be operationally feasible, meaning they could be executed step-by-step to achieve the privilege escalation goal. This output of multiple, non-identical chains provides security analysts with a broader perspective on potential attack vectors, moving beyond a single "ground-truth" path to a more comprehensive risk assessment. Finally, we provide a brief description and analysis of each attack chain in Table~\ref{tab:ctf_vm}.

\begin{table*}[tbp]
\centering
\caption{Performance on CTF Virtual Machines.}
\label{tab:ctf_vm}
\begin{tabular}{cccccl}
\hline
\textbf{CTF VM}  & \textbf{Success} & \textbf{\begin{tabular}[c]{@{}c@{}}\# Chains \\ Found\end{tabular}} & \textbf{TTFS (s)} & \textbf{\begin{tabular}[c]{@{}c@{}}Shortest\\ Length\end{tabular}} & \textbf{Chain Description and Analysis}                                                                                                                                                                                       \\ \hline
C1               & Yes              & 3                                                                   & 593               & 6                                                                  & \begin{tabular}[c]{@{}l@{}}Chains involved three primitives: SUID binary exploitation, \\ cron job hijacking, and writable systemd service corruption.\end{tabular}                                                           \\
C2               & Yes              & 2                                                                   & 835               & 5                                                                  & \begin{tabular}[c]{@{}l@{}}Two distinct vectors discovered: .bashrc environment variable\\  hijack and direct systemd service file corruption.\end{tabular}                                                                   \\
C3               & Yes              & 4                                                                   & 816               & 8                                                                  & \begin{tabular}[c]{@{}l@{}}Multiple chains identified combining LD\_PRELOAD injection, \\ sudo misconfiguration exploitation, and \\ symbolic link attacks.\end{tabular}                                                      \\
C4               & Yes              & 2                                                                   & 692               & 7                                                                  & \begin{tabular}[c]{@{}l@{}}Alternative paths through different writable script targets: \\ Python service script manipulation and shell script hijacking.\end{tabular}                                                        \\
C5               & Yes              & 1                                                                   & 1417              & 9                                                                  & \begin{tabular}[c]{@{}l@{}}Exploit the command injection vulnerability in the scanlosers.sh \\ script to gain access as the \texttt{pwn} user and leverage passwordless \\ sudo privileges to execute msfconsole as root.\end{tabular} \\ \hline
\textbf{Average} & \textbf{100\%}   & \textbf{2.4}                                                        & \textbf{870.6}    & \textbf{7.0}                                                       & All 12 total chains were manually verified as feasible.                                                                                                                                                                       \\ \hline
\end{tabular}
\end{table*}

Analysis for RQ1 \& RQ3: These results confirm \sn{}'s effectiveness in discovering viable attack chains within operationally relevant timeframes (RQ1). More importantly, the consistent generation of multiple distinct and validated chains – rather than converging on a single optimal solution – directly addresses the solution inflexibility challenge. This capability provides defenders with a more complete understanding of the attack surface, enabling them to prioritize mitigations that break entire classes of attacks rather than just the most obvious path.

\par
\begin{figure}[h!t]
\centering
\includegraphics[width=\linewidth]{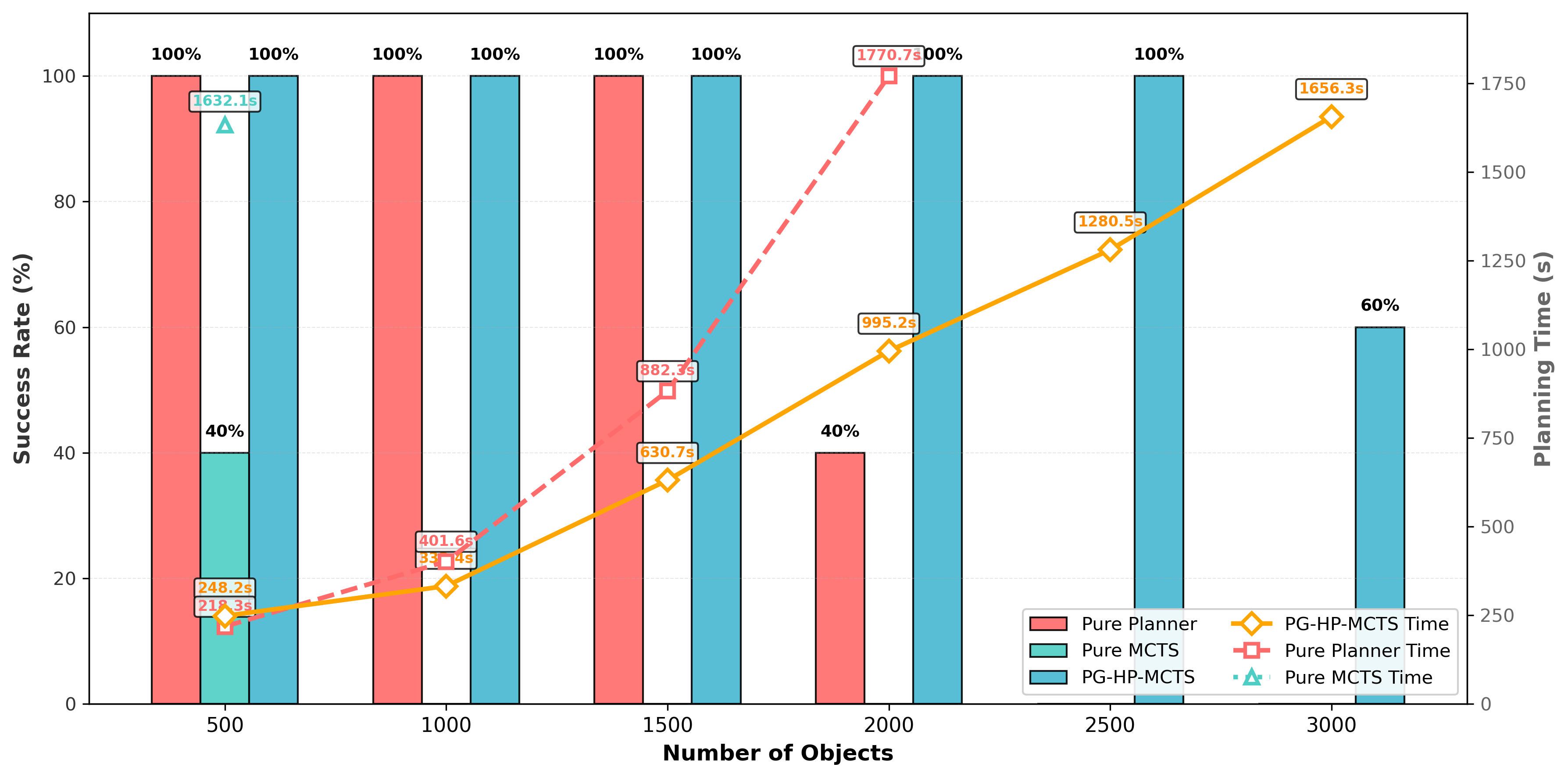}
\caption{Efficiency and Scalability Analysis: Success Rate \& Planning Time.}
\label{fig:efficiency_scalability}
\end{figure}

\textbf{Synthetic Problems: Efficiency and Scalability.}Figure~\ref{fig:efficiency_scalability} and Table~\ref{tab:syn_problems} collectively demonstrate the high efficiency and scalability of \sn{} compared to baseline approaches on synthetic datasets. The average success rates for the pure optimal planner, pure MCTS, and \sn{} were 57\%(17), 7\%(2), and 93\%(28), respectively. Notably, the pure optimal planner and pure MCTS achieved zero solution success rates at two (2500, 3000) and five (1000, 1500, 2000, 2500, 3000) scales, respectively. In contrast, \sn{} demonstrates exceptional efficiency and scalability, achieving a 60\% planning success rate even on the largest-scale problem (3000 objects)—where competing approaches both failed completely. 
We additionally focused on analyzing the primary reasons for the poor performance of pure MCTS. As shown in Table~\ref{tab:syn_problems}, the average number of node branches and action enumeration time for pure MCTS were significantly higher than those for PG-HP-MCTS. This is primarily because pure MCTS lacks both phase guidance and efficient enumeration capabilities. Notably, the action enumeration process in pure MCTS consumed 91\% of the total solution time.
Finally, we have calculated that the phase-guided PG-HP-MCTS exhibits an average behavioral enumeration count of 219. This represents only 23\% of the average enumeration count of 952 observed in the absence of phase guidance.
\par
Analysis for RQ2: \sn{}'s significantly higher success rate and lower planning time, especially on complex problems, provide strong evidence that the PG-HP-MCTS algorithm successfully tackles the computational inefficiency challenge. The drastic reduction in states expanded indicates effective pruning via phase guidance.

\begin{table}[tbp]
\centering
\caption{Aggregate Results on Synthetic Problems (n=30).}
\label{tab:syn_problems}
\begin{tabular}{cccc}
\hline
\textbf{Metric}                                                               & \textbf{\begin{tabular}[c]{@{}c@{}}Pure \\ Planner\end{tabular}} & \textbf{\begin{tabular}[c]{@{}c@{}}Pure \\ MCTS\end{tabular}} & \textbf{PG-HP-MCTS} \\ \hline
\textbf{Success Rate}                                                         & 57\%                                                             & 7\%                                                           & \textbf{93\%}       \\
\textbf{\begin{tabular}[c]{@{}c@{}}Avg. Planning \\ Time (s)\end{tabular}}    & \textbf{649.5}                                                   & 1632.7                                                        & 799.8               \\
\textbf{\begin{tabular}[c]{@{}c@{}}Avg. Branches\\ Per Node\end{tabular}}     & /                                                                & 27.3                                                          & \textbf{8.1}        \\
\textbf{\begin{tabular}[c]{@{}c@{}}Avg. Action\\ Enum. Time (s)\end{tabular}} & /                                                                & 104.5                                                         & \textbf{5.1}        \\ \hline
\end{tabular}
\end{table}

\subsection{Large-Scale Real-World Evaluation}
We deployed \sn{} on 291 live cloud instances to evaluate its practical effectiveness in real-world scenarios. The evaluation targeted 175 Amazon EC2 instances and 116 Alibaba Cloud ECS instances, representing a diverse set of official and marketplace Linux distributions. The information collection phase demonstrated efficiency, requiring an average of 3.1 minutes per instance with a maximum of 5.5 minutes.
\par
In this large-scale deployment, \sn{} successfully discovered privilege escalation chains in 27 instances, comprising 19 EC2 instances (10.9\%) and 8 ECS instances (6.9\%). More significantly, the system identified 9 previously unreported attack chains that represent novel security vulnerabilities in production cloud environments. The average total analysis time—including both information collection and solving phases—was approximately 19.4 minutes per instance, representing a reduction of over 47.6\% compared to ChainReactor's reported average of 37 minutes~\cite{de2024chainreactor}. This efficiency gain directly stems from our PG-HP-MCTS algorithm's optimized search strategy.
\par
The discovered attack chains demonstrated meaningful diversity and sophistication. On average, each vulnerable instance yielded 2.1 distinct attack paths, with EC2 instances showing slightly higher diversity (2.3 chains) than ECS instances (1.8 chains). This multi-solution capability proved particularly valuable for defensive analysis, as it revealed alternative attack vectors that might otherwise remain hidden when using single-solution approaches.
\par
Case Study 1 (Docker Group to Service File Corruption Chain): On a \texttt{Debian-11} EC2 instance, \sn{} discovered a 4-step chain based purely on permission misconfigurations. The chain began with the unprivileged user being a member of the \texttt{docker} group, granting container management capabilities. Further analysis revealed a systemd service (\path{/etc/systemd/system/metrics-collector.service}) owned by the \texttt{docker} group with group-write permissions. Using Docker's ability to mount the host filesystem, the attacker could modify this service file to execute arbitrary commands. Since the service was configured to start at boot with root privileges, the modification resulted in privilege escalation.
\par
Case Study 2 (SUID Binary to Cron Script Hijacking): On an \texttt{Ubuntu-20.04} ECS instance, the system uncovered a chain combining SUID binary misuse with cron script manipulation. The initial foothold involved a custom SUID binary (\texttt{/usr/local/bin/backup-tool}) that executed shell commands without proper sanitization. While this binary alone did not provide direct root access, \sn{} discovered it could be used to modify a world-writable cron script (\texttt{/etc/cron.daily/disk-cleanup}) that ran as root. By chaining these two misconfigurations—first using the SUID binary to inject malicious commands into the cron script, then relying on the cron execution—the system achieved privilege escalation. 
\par
Case Study 3 (Environment Variable Injection via Writable Profile Scripts): On a \texttt{Centos-7 }EC2 instance, \sn{} identified a subtle chain involving environment variable manipulation. The system detected that the user's home directory contained a \texttt{.profile} script that was writable by other users in the same group. Concurrently, a system administration script (\texttt{/usr/local/bin/system-check}) executed with root privileges and sourced environment variables without proper validation. By modifying the \texttt{.profile} script to inject malicious environment variables, and triggering the system-check script (which could be done through various means discovered by the solver), the attacker could execute arbitrary commands as root. 
\par
Analysis for RQ1 \& RQ4: The discovery of 9 unreported attack chains in production cloud environments provides compelling evidence of \sn{}'s practical effectiveness (RQ1). More importantly, the consistent generation of multiple attack chains per vulnerable instance—combined with significantly reduced analysis times—demonstrates that our approach not only identifies novel vulnerabilities but does so with efficiency that enables large-scale security assessments (RQ4). The 47.6\% reduction in analysis time compared to ChainReactor directly addresses operational constraints in real-world security workflows, while the multi-chain output provides defenders with a more comprehensive understanding of their attack surface. Notably, all discovered chains involved either pure configuration errors or hybrid chains combining vulnerabilities with misconfigurations, validating our domain modeling approach focused on system interactions rather than individual bug exploitation.
\par
\textbf{Direct Comparison with ChainReactor.} To ensure a fair comparison, we ran the reproduced ChainReactor pipeline on a randomly selected subset of 50 instances from our dataset.  It is important to note that ChainReactor directly invokes pure solver (Powerlifted) for its solving component. As shown in Table~\ref{tab:compare_chainreactor}, the results highlight the differences between ChainReactor (a pure satisfying solver) and our method. First, \sn{} identified potential attack chains on 6 instances, while ChainReactor found only 2 (which were covered), indicating that \sn{} possesses stronger attack chain detection sensitivity. Second, \sn{} discovered a total of 14 attack chains—seven times more than ChainReactor—with an average of 2.3 attack chains per instance, which is 2.3 times higher than ChainReactor's average. Finally, \sn{} achieved an average solution time of 14.4 minutes, representing a 42.6\% reduction in time overhead compared to ChainReactor. 
\par
This direct comparison validates that our PG-HP-MCTS algorithm is not only faster but crucially generates a richer set of offensive options. Where ChainReactor stops at the first found path, \sn{} continues exploring, which is vital for comprehensive defensive analysis (RQ3). The speed advantage stems from the targeted, phase-guided search (RQ2).

\begin{table}[tbp]
\centering
\caption{Direct Comparison with ChainReactor.}
\label{tab:compare_chainreactor}
\begin{tabular}{lll}
\hline
\textbf{Metric}                        & ChainReactor & \sn{}  \\ \hline
\textbf{Instances with $\ge$1 Chain}    & 2            & 6      \\
\textbf{Total Chains Discovered}       & 2            & 14     \\
\textbf{Avg. Chains per Instance}      & 1.0          & 2.3    \\
\textbf{Avg. Solver Time per Instance} & 25.1min      & 14.4min \\ \hline
\end{tabular}
\end{table}

\section{Discussion \& Limitation}
\label{sec:discussion}

\subsection{Discussion}
\label{subsec:discussion}
Our experimental results demonstrate that \sn{} effectively addresses core challenges in automated attack chain exploration. The phase-guided hybrid algorithm successfully balances computational efficiency with solution diversity, enabling discovery of multiple feasible chains within operationally relevant timeframes. This capability represents a significant advancement over traditional satisficing planners, which terminate upon finding a single solution.
\par
The generation of diverse attack chains has particular defensive value. Rather than identifying only the shortest path, \sn{} reveals different vulnerability classes and their interactions within a system. This comprehensive view enables defenders to prioritize mitigations that break entire attack surfaces—for instance, fixing a single writable service file that appears in multiple chains—rather than addressing isolated vulnerabilities. Our approach thus shifts the defensive paradigm from reactive patching to proactive attack surface reduction.
\par
The methodology is fundamentally extensible. While instantiated for Unix post-exploitation, the core framework—formal domain modeling combined with phase-guided hybrid search—can be adapted to other security domains requiring multi-step reasoning, such as cloud misconfiguration analysis or software supply chain attack exploration. The primary adaptation would involve creating a new domain model and designing phase-guided hybrid MCTS algorithms; the solver architecture remains largely domain-agnostic.

\subsection{Limitations and Future Work}
\label{subsec:limitations}
Despite its contributions, \sn{} has several limitations that warrant discussion. First, while our domain model captures prevalent attack vectors, it remains an abstraction. Certain advanced techniques—particularly kernel-level exploits, race conditions, or hardware-based attacks—fall outside its current scope. Future work will develop a modular extension framework to incorporate new attack patterns as they emerge, potentially through automated extraction from security advisories and exploit write-ups.
\par
Second, our current vulnerability capability modeling requires manual curation. We manually extracted specific capabilities (e.g., arbitrary file write, command injection) from known CVEs and mapped them to executable predicates. This process is labor-intensive and cannot scale to the continuous stream of new vulnerabilities. Future work should investigate automated vulnerability capability extraction using large language models (LLMs). By analyzing CVE descriptions, exploit proofs-of-concept, and advisory texts, LLMs could automatically infer the security-relevant effects of vulnerabilities and translate them into formal planning predicates and action schemas. This would significantly enhance the model's comprehensiveness and timeliness.
\par
Third, our current model optimizes for logical feasibility rather than operational stealth. Generated chains may contain actions likely to trigger security alerts in monitored environments. Integrating detection-awareness into the reward function—by modeling common EDR signatures or logging patterns—would enable the discovery of more covert attack paths, better simulating advanced adversaries.
\par
Fourth, \sn{} generates attack chains but does not automatically execute them to validate operational feasibility or collect runtime data. Future work should implement safe, controlled execution of discovered chains in isolated environments, coupled with comprehensive logging of system events. This would enable the creation of high-quality, actionable datasets for data-driven network/host security. Such datasets, capturing the complex multi-step nature of real attacks, could significantly improve supervised machine learning approaches for intrusion detection that currently struggle with sophisticated, multi-vector threats.
\par
Finally, \sn{} assumes static host configurations during analysis and does not model interactive defender responses. Extending the framework to support adaptive planning in dynamic environments, possibly through integration with lightweight runtime monitors, represents a valuable direction for transforming attack chain exploration into fully automated penetration testing.

\section{Related Work}
\label{sec:relatedwork}
\subsection{Automated Penetration Testing and Post-Exploitation Analysis}
The automation of security assessment has evolved along two primary dimensions: pre-exploitation vulnerability discovery and post-exploitation attack chain analysis. In the pre-exploitation domain, systems like PenTestGPT and PentestAgent~\cite{deng2024pentestgpt, shen2025pentestagent} leverage large language models to interpret scan results and generate exploitation scripts, while some approaches~\cite{zhou2025mind, schwartz2019autonomous, hu2020automated} employ reinforcement learning to explore networks and establish footholds. These approaches focus on the early stages of the attack lifecycle—reconnaissance, vulnerability scanning, and initial compromise. Complementary to these efforts are traditional vulnerability discovery techniques including fuzzing~\cite{fioraldi2020afl++-17, liang2018fuzzing-37}, symbolic execution~\cite{baldoni2018survey-4}, hybrid analysis~\cite{chen2022symsan-7, poeplau2020symbolic-55, stephens2016driller-69}, bounded model checking~\cite{biere2021bounded-5}, static analysis~\cite{chess2004static-8}, and formal verification~\cite{hasan2015formal-25}. While effective for identifying individual vulnerabilities, these methods lack the capability to reason about how multiple weaknesses can be strategically combined into operational attack chains.

In contrast, our work focuses exclusively on the post-exploitation phase, where an attacker has already obtained shell access and must navigate complex host environments. This requires fundamentally different modeling that captures system state transitions, permission relationships, file dependencies, and capability propagation—elements that are largely irrelevant to initial access tools but essential for discovering multi-step attack chains that blend benign and malicious operations. Our approach bridges this gap by providing systematic methodology for discovering how defenders can proactively identify chains combining multiple independent weaknesses.

\subsection{Formal Modeling and Planning for Attack Simulation}
Formal methods have been widely applied to model security operations and enable systematic reasoning. Early simulation approaches modeled attacks as cause-effect transitions~\cite{cohen1999simulating}, while graph-based representations like attack graphs have been used to analyze network intrusion paths~\cite{holm2014p, johnson2018meta, ou2005mulval}. Other formalisms include Petri nets~\cite{pudar2009penet, zhang2016attack}, logical frameworks~\cite{kotenko2003active}, and game-theoretic models~\cite{abercrombie2014security}. More recently, AI planning has shown particular promise for modeling multi-step attacks, with applications in network penetration testing~\cite{amos2017efficient-3, hoffmann2015simulated-27, miller2018automated-39, speicher2019towards-68} and smart grid security~\cite{choi2021plan2defend-9}.

While planning has also been applied to host-level privilege escalation~\cite{de2024chainreactor, maeda2021automating}, existing approaches face significant challenges when modeling complex host environments. Most critically, they typically employ satisficing strategies that terminate upon finding any feasible solution, providing limited insight into alternative attack paths. This limitation is particularly problematic in post-exploitation scenarios where defenders need to understand all potential attack vectors, not just the most obvious one. Our work addresses this by integrating planning with Monte Carlo Tree Search (MCTS), enabling discovery of multiple diverse attack chains while maintaining the formal correctness guarantees of classical planning.

\subsection{Search Algorithms and System-Specific Security Analysis}
Exploring the vast solution space of possible attack chains requires efficient search algorithms capable of handling complex state spaces. Classical planning systems employ either grounded~\cite{correa2023grounding-13, gnad2019learning-21, helmert2009concise-26} or lifted~\cite{correa2020lifted-14, ridder2014lifted-62} approaches, with lifted planners offering better scalability for domains with many objects—a critical requirement for host modeling. Monte Carlo Tree Search provides complementary strengths through its ability to balance exploration and exploitation via randomized simulations~\cite{kocsis2006bandit}, offering anytime performance and multi-solution discovery. However, pure MCTS lacks the formal reasoning needed for complex security operations.

Our phase-guided hybrid Planning-MCTS algorithm synthesizes these approaches, addressing limitations in both paradigms. Unlike penetration testing tools that explore abstract network spaces~\cite{kong2025vulnbot, li2025her}, our algorithm is specifically designed for the concrete, highly constrained environments of post-exploitation host analysis. This domain requires reasoning about specific system components and their complex interactions, rather than abstract network states.

The problem of privilege escalation has been extensively studied across various environments, including UNIX/Linux systems~\cite{jain2014practical-31, jeon2019polper-32}, mobile platforms~\cite{aldoseri2022tale-2, davi2010privilege-15}, browser environments~\cite{kim2023extending-35}, trusted execution environments~\cite{suciu2020horizontal-70}, cloud infrastructures~\cite{haimed2023exploiting-24, pecka2022privilege-52}, and firmware~\cite{yin2022finding-77}. While these investigations comprehensively catalog individual escalation mechanisms, they provide limited methodology for automatically discovering how these mechanisms can be chained together—the precise gap our work addresses.

\subsection{Positioning of Our Work}
\textsc{ChainReactor}~\cite{de2024chainreactor} represents the closest related work, applying lifted planning to privilege escalation chain discovery. Our approach differs fundamentally in several aspects. First, we introduce semantic phase guidance that organizes attack actions based on their operational purpose, enabling more efficient search. Second, we integrate MCTS with planning rather than relying solely on classical planners, addressing solution inflexibility and enabling discovery of multiple diverse chains. Third, our modeling captures finer-grained host state transitions and attacker capability evolution during post-exploitation. Finally, our evaluation demonstrates discovery of previously unreported attack chains in production cloud environments.

\section{Conclusion}
\label{sec:conclusion}
In this work, we presented \sn{}, an automated system for discovering post-exploitation attack chains in complex host environments. \sn{} addresses the realistic threat scenario where adversaries combine benign operations, configuration errors, and targeted exploits to form multi-step attack chains. By formulating this problem as an AI planning task and solving it with our novel phase-guided hybrid Planning-MCTS algorithm, we enable efficient and diverse exploration of attack chains that would be infeasible to discover manually.

Our experimental evaluation demonstrates the practical effectiveness of this approach. \sn{} successfully rediscovers known attack chains in controlled benchmarks while identifying 9 previously unreported privilege escalation vectors in real-world cloud environments. The system's ability to generate multiple distinct attack paths provides defenders with comprehensive threat intelligence, enabling proactive hardening of complex production systems against sophisticated post-exploitation attacks.

\bibliographystyle{unsrt}

\bibliography{ref.bib}

\clearpage
\appendices
\section{Extended Discussion on Post-Exploitation and MITRE ATT\&CK}
\label{app:mitre-attack-details}
Within the MITRE ATT\&CK Matrix~\cite{attck_matrix}, 11 of the 14 tactics fall under the post-exploitation phase (excluding reconnaissance, resource development, and initial access). In contemporary attack scenarios, particularly Advanced Persistent Threats (APTs), post-exploitation has evolved into a sophisticated process combining both malicious exploits and benign system interactions. Attackers typically operate under constrained privileges initially and must carefully orchestrate multi-step attack chains to reach their goals. These chains often leverage system misconfigurations, insecure permissions, vulnerable software versions, and legitimate system utilities in creative combinations. For instance, attackers might abuse sudo misconfigurations, hijack shared library loading mechanisms, exploit cron job vulnerabilities, or manipulate environment variables to escalate privileges.
\par
The complexity of modern operating systems creates numerous potential attack vectors during post-exploitation. File system permissions, user-group relationships, process capabilities, scheduled tasks, and service configurations all present opportunities for privilege escalation and persistence. What makes post-exploitation particularly challenging for defenders is that these attack chains often blend malicious actions with legitimate system operations, making detection through conventional security mechanisms difficult. Understanding these complex interactions requires a systematic analysis of the host environment and the potential pathways an attacker could leverage to advance their position.

\section{Formal Planning Definitions and PDDL Specification}
\label{app:planning-formalization}
\subsection{Formal Planning Problem}
A classical planning problem is typically defined by the tuple $<S, A, \gamma, s_{o}, S_{G}>$, where $S$ represents the set of all possible states, $A$ is the set of available actions, $\gamma: S\times A\to S$ is the state transition function, $s_{o} \in  S$ is the initial state, and $S_{G} \in  S$ is the set of goal states. A solution to such a problem is a sequence of actions $<a_{1}, a_{2}, ..., a_{n}>$ that, when applied in sequence starting from $s_{o}$, results in a state $s_{n}\in S_{G}$.
\subsection{PDDL Domain and Problem Specification}
The Planning Domain Definition Language (PDDL)~\cite{aeronautiques1998pddl} provides a standardized formalism for describing planning problems through domain and problem files. The domain file defines the static structure using objects (e.g., users, groups, different types of files, and file system permissions), predicates representing system facts (e.g., file capabilities, content relationships), and actions modeling state transitions. Each action specifies parameters, preconditions (conditions required for execution), and effects (fact changes after execution). In our work, the domain file (domain.pddl) is heuristically defined based on our analysis of post-exploitation behaviors, formally capturing the objects, facts, and actions relevant to host-level attacks.
\par
Predicates describe the facts that hold in a given state. In PDDL syntax, a predicate takes the form \texttt{(predicate\_name ?arg1 - type1 ?arg2 - type2 ...)} where \texttt{?arg} denotes a variable (parameter) of the specified type. For instance, \texttt{(user\_write\_access ?u - user ?g - group ?f - file)} indicates that user \texttt{?u} in group \texttt{?g} has write access to file \texttt{?f}. When all variables are instantiated with concrete objects (e.g., \texttt{(user\_has\_access\_write alice developers config.txt)}), the predicate becomes a grounded fact that is either true or false in a particular state.
\par
Actions model state transitions with the following components: (1) Parameters (\texttt{?var - type}): Variables that must be instantiated with concrete objects; (2) Precondition: A logical formula that must hold true in the current state for the action to be applicable. (3)Effect: Changes to the state when the action is executed, adding (\texttt{and (predicate ...)}) or removing (\texttt{and (not (predicate ...))}) facts.
\par
The problem file instantiates the domain with concrete objects (specific files, users, etc.), defines the initial state through grounded predicates describing the current system configuration, and specifies the goal condition representing the attacker's objective. \sn{} automatically generates this problem file by retrieving the state of the target host, including the file system, the relationships between user groups, and the executable properties, during the post-exploitation phase with limited privileges.

\section{Extended Motivation Example}
\label{app:motivation-example}
This appendix provides the detailed example referenced in the Motivation \& Threat Model section of the main manuscript to illustrate the complexity of modern post-exploitation attack chains.
\par
Consider a privilege escalation scenario on a Linux system, where an attacker operates solely with the initial privileges of a low-privileged user (\texttt{user\_a}):
\begin{enumerate}
    \item \textbf{Benign Enumeration}: The attacker begins by understanding the automated processes on the system. They list the cron jobs and discover that user \texttt{user\_b} has a scheduled task that periodically executes a backup script located at \texttt{/home/user\_b/scripts/backup.sh}. Crucially, this script is configured to run with root privileges via sudo without a password.
    
    \item \textbf{Vulnerability Discovery}: The attacker inventories the system for vulnerable software. A critical finding emerges: the system's gzip binary (\texttt{/bin/gzip}) is version 1.11. This version is known to be affected by CVE-2022-1271, an arbitrary file write vulnerability that allows attackers to write arbitrary content to any file location accessible by the user running \texttt{gzip}.
    
    \item \textbf{Malicious Orchestration}: Leveraging this vulnerability, the attacker crafts an exploit. They use the \texttt{gzip -N} command with specially crafted input to directly append malicious shell commands to the privileged script \texttt{backup.sh}. The exploit carefully constructs the payload to ensure the appended content constitutes valid shell commands that will execute during the script's next run.
    
    \item \textbf{Benign Trigger}: The attacker waits for the scheduled cron job to execute. The cron daemon runs \texttt{/home/user\_b/scripts/backup.sh} with root privileges. The script now executes both its legitimate backup operations and the malicious payload appended by the attacker.
    
    \item \textbf{Privilege Escalation}: The malicious payload executes in the context of the root-privileged process, immediately granting the attacker a root shell on the system.
\end{enumerate}

This attack chain exemplifies the core challenge in modern post-exploitation: successful attacks are rarely about a single vulnerability. Instead, they are sophisticated sequences that strategically chain together system misconfigurations, version-specific software vulnerabilities, and benign and malicious actions. It is the intricate interplay and specific combination of these distinct elements—each potentially harmless on its own—that creates a viable path to a critical security objective.

\section{Design Principles Detailed Explanation}
\label{app:principles-details}

\subsection{Formal Abstraction Principle}
We identify the minimal yet sufficient set of PDDL constructs to capture post-exploitation mechanics, ensuring all reasoning is logically sound, reproducible, and free from ambiguities of natural language or ad-hoc scripting.

\subsection{Phase-Guided Search Principle}
By observing that post-exploitation actions naturally progress through semantically meaningful phases, we incorporate this structure into the search strategy. Phase-specific constraints and heuristics prune irrelevant branches early, dramatically improving computational efficiency and focusing exploration on promising avenues.

\subsection{Hybrid Solving Principle}
Our approach synergistically combines symbolic AI planning with probabilistic search. We leverage planners for well-defined sub-problems within phases while using MCTS for high-level strategy management, achieving both rigor in constraint satisfaction and exploratory power in large state spaces.

\subsection{Minimal Interference Principle}
During information collection, \sn{} operates under strict read-only mandates: no file creation/modification/deletion, no service/process manipulation, and no configuration alterations. This ensures passive, forensically quiet assessment safe for production environments.

\section{Modeling Philosophy Details}
\label{app:modeling-details}
\subsection{Attack Surface Details}
For each of the seven attack surfaces:
\begin{itemize}
    \item File system permissions and ownership: The foundational layer of Unix security. Without modeling permissions, we cannot reason about whether an attacker can read, write, or execute files—the basic building blocks of any attack chain.
    \item Process execution and privilege inheritance: Processes are the vehicles through which attackers execute code. Modeling process creation, ownership, and SUID/SGID behavior is essential for capturing privilege escalation.
    \item Automated execution mechanisms: Cron jobs, systemd services, and other scheduled tasks represent critical pivots—they execute with specific privileges at predictable times, making them prime targets for persistence and privilege escalation.
    \item Dependency relationships: Shared libraries, configuration files, and script inclusions create implicit trust relationships. An attacker controlling a loaded file can influence the loader's behavior—a common exploitation pattern.
    \item Software vulnerabilities: Known CVEs provide capabilities that bypass normal permission checks. Integrating CVE knowledge allows the model to capture real-world exploit chains combining misconfigurations with vulnerabilities.
    \item User and group relationships: Unix privilege models are fundamentally based on user and group memberships. Modeling these relationships is necessary for understanding permission inheritance and group-based privilege escalation.
    \item Data movement capabilities: File upload and download represent basic attacker capabilities for payload delivery and data exfiltration.
\end{itemize}
\subsection{Omitted Aspects}
This model deliberately omits: vulnerability exploitation process, network stack states, specific memory layouts, kernel internals, and fine-grained process scheduling. These are not required for reasoning about the logical dependency chains of file permissions, user contexts, and automated executions that define host-level post-exploitation.

\section{PDDL Domain Details}
\label{app:pddl_details}
\textbf{Predicates (State Properties)}. 
Predicates describe the facts that hold in a given state. In PDDL syntax, a predicate takes the form \texttt{(predicate\_name ?arg1 - type1 ?arg2 - type2 ...)} where \texttt{?arg} denotes a variable (parameter) of the specified type. For instance, \texttt{(user\_write\_access ?u - user ?g - group ?f - file)} indicates that user \texttt{?u} in group \texttt{?g} has write access to file \texttt{?f}. When all variables are instantiated with concrete objects (e.g., \texttt{(user\_has\_access\_write alice developers config.txt)}), the predicate becomes a grounded fact that is either true or false in a particular state.
We define a comprehensive set of 30 predicates and categorize them into five functional groups:
\par
(1) Capability Predicates (11 predicates) describe what executables can perform, either through native functionality or known vulnerabilities. This includes basic operations like file manipulation (\texttt{can\_read/write/upload/download\_file}, \texttt{can\_change\_file\_onwer}, \texttt{can\_change\_file\_permission}), system interaction capabilities (\texttt{can\_exec\_cmd}, \texttt{can\_shell}), and vulnerability-based predicates (\texttt{has\_cve\_cmd\_inject}, \texttt{has\_cve\_any\_write/read}) that model CVE exploits bypassing normal permission checks. These predicates abstract implementation details, focusing on the effects executables can achieve.
\par
(2) Privilege and Control Predicates (4 predicates) track the attacker's progress and system privilege structure. Key predicates include \texttt{attacker\_control} (indicating which users are under attacker control), \texttt{user\_is\_admin} (administrative privilege status), \texttt{user\_in\_group} (membership relationships), and \texttt{group\_is\_admin} (group-level administrative privileges).
\par
(3) File System Predicates (11 predicates) capture static properties and dynamic state of the file system. These include permission matrices (\texttt{user/group\_has\_file\_perm}), ownership (\texttt{file\_belongs\_to}), content storage (\texttt{file\_contains}), and special file properties enabling specific attack vectors: \texttt{has\_suid\_flag} (SUID binaries), \texttt{scheduled\_exec\_by\_user} (cron/scheduled tasks), \texttt{exec\_auto\_loads\_file} (auto-loaded dependencies), and \texttt{is\_daemon\_file} (service management files). State tracking predicates (\texttt{file\_was\_read}, \texttt{file\_was\_uploaded}, \texttt{file\_was\_downloaded}) record file access for exfiltration analysis.
\par
(4) Process Execution Predicate (1 predicate) \texttt{process\_running} models runtime state by serving as an execution token: it represents that process \texttt{?p} is running with user \texttt{?u} executing binary \texttt{?e}. This design enforces realistic sequencing—actions requiring active process contexts consume this token, while process creation actions generate it.
\par
(5) Derived Capability Predicates (3 predicates) compute effective permissions from base predicates (\texttt{user\_has\_read/write/exec\_access}). These are not directly asserted but derived through axiom actions that consolidate all permission sources: direct user permissions, group permissions, ownership, and administrative overrides.
\par
\textbf{Actions (State Transitions)}. 
The domain includes 19 actions modeling state transitions available to the attacker, categorized by their primary purpose:
\par
(1) Process Execution Actions (3 actions) establish and leverage execution contexts. \texttt{launch\_process} creates basic process tokens, while \texttt{launch\_suid\_process} enables SUID-based privilege escalation—if an attacker-controlled user can execute a SUID binary, they obtain a process running as the binary's owner. \texttt{launch\_shell\_grant\_control} directly grants user control when a shell-capable process is executed.
\par
(2) File Manipulation Actions (7 actions) modify file system state. \texttt{write\_file} allows writing data to writable files using appropriate executables, while \texttt{cve\_write\_any\_file} bypasses permission checks via arbitrary-write vulnerabilities. \texttt{read\_file} and  \texttt{cve\_read\_any\_file}captures file reading for information gathering, and \texttt{propagate\_loaded\_content} models dependency chain exploitation—if an executable always loads a file, controlling that file's contents effectively controls the loader (e.g., .bashrc hijacking). \texttt{download\_file} creates attacker-owned files locally, while \texttt{upload\_file} marks files as exfiltrated to remote systems.
\par
(3) System Configuration Actions (4 actions) alter access controls and system state. \texttt{set\_suid\_flag} allows setting SUID bits (post-escalation backdoor creation), \texttt{change\_file\_owner} modifies ownership, and \texttt{add\_default\_perm} adds default permissions to owned files. \texttt{hijack\_scheduled\_exec} exploits scheduled tasks: writing a malicious payload to an executable systematically called by another user grants control of that user when the task executes.
\par
(4) Derivation and Capability Propagation Actions (5 actions) automatically infer facts when preconditions are met. These include permission derivation axioms (\texttt{derive\_read/write/exec\_access}) that compute effective access rights from multiple permission sources. Capability propagation actions model logical relationships: \texttt{derive\_cmd\_implies\_caps} establishes that arbitrary command execution implies multiple other capabilities, while \texttt{derive\_launch\_shell} transforms payloads into a shell-spawning capability.

\section{Script Implementation Details}
\label{app:scripts-details}
The information extraction is organized into five core categories, each fulfilling specific requirements of the modeling philosophy and domain definition:
\par
(1) Identity and Privilege Context – This corresponds to the user and group relationships attack surface identified in the modeling philosophy. The scripts enumerate all system users and groups from \texttt{/etc/passwd} and \texttt{/etc/group}, establishing the concrete user and group objects for the domain. We further parse membership relationships to instantiate the \texttt{user\_in\_group} predicate and identify the current execution context to initialize the \texttt{attacker\_control} predicate for the foothold user.
\par
(2) File System State – This addresses the file system permissions, ownership, and process execution and privilege inheritance attack surfaces. The collection phase discovers all files in critical directories (\texttt{/bin}, \texttt{/usr/bin}, \texttt{/home}, etc.), classifies them into the domain's type hierarchy (executable, sharedfile, regular file), and extracts their security metadata. This includes octal permissions (grounding the \texttt{user\_has\_file\_perm} and \texttt{group\_has\_file\_perm} predicates), ownership information (for \texttt{\texttt{file\_belongs\_to}}), and special permission flags like SUID/SGID (for \texttt{has\_suid\_flag}).
\par
(3) System Dynamics – This operationalizes the automated execution mechanisms and dependency relationships components of the modeling philosophy. The scripts identify all scheduled execution mechanisms: cron jobs (from \texttt{/etc/crontab}) and systemd services (from \texttt{/etc/systemd/}), creating calledexecutable objects and asserting the \texttt{scheduled\_exec\_by\_user} predicate. We also map dependency chains by analyzing shared library loads and shell configuration files (e.g., \texttt{.bashrc}, \texttt{.profile}), establishing facts for the \texttt{exec\_auto\_loads\_file} predicate that enables modeling of library hijacking and shell initialization attacks.
\par
(4) Software Vulnerability Context – This implements the software vulnerabilities dimension of the attack surface model. The scripts query version information for discovered executables and validate them against a knowledge base of known CVEs. When an executable matches a vulnerable version pattern and satisfies all dependency conditions, the corresponding CVE capability predicate (e.g., \texttt{has\_cve\_any\_write}, \texttt{has\_cve\_cmd\_inject}) is asserted in the problem state, enabling the planner to reason about exploit chains.
\par
(5) Capability Derivation Primitives – Supporting the data movement capabilities and the efficient permission derivation strategy outlined in the domain model, the scripts identify writable files and directories not owned by the current user. This information, while not directly mapping to a single predicate, provides the raw data from which the derived capability predicates (\texttt{user\_has\_access\_read/write/exec}) can be computed by the planner's axiom actions, avoiding the storage of a quadratic number of permission facts.

\section{Report Generation Details}
\label{app:report-details}

\textbf{LLM Prompt Template and Example.} Our LLM prompt transforms filtered formal PDDL attack chains into actionable natural language reports.

\subsection{Input: PDDL Attack Chain Example (Docker Group Escalation)}
\begingroup\small
\begin{verbatim}
1: (launch_process ec2-user_u docker_g 
usr_bin_docker process1)
2: (write_file process1 
usr_bin_docker 
etc_systemd_system_cloud-logger_timer 
shell_payload local ec2-user_u 
docker_g)
3: (spawn_shell_grant_control
etc_systemd_system_cloud-logger_timer 
root_u process2)
\end{verbatim}
\endgroup

\textbf{PDDL Object Explanations}:
\begin{itemize}
    \item \texttt{ec2-user\_u}: The default EC2 user account.
    \item \texttt{docker\_g}: Docker group.
    \item \texttt{etc\_systemd\_system\_cloud-logger\_timer}: Systemd timer file scheduled for execution by \texttt{ec2-user}.
    \item \texttt{usr\_bin\_docker}: Docker binary at \texttt{/usr/bin/docker}.
    \item \texttt{root\_u}: Root user account.
\end{itemize}

\subsection{LLM Prompt Template}
\begingroup\small
\begin{verbatim}
Analyze this PDDL attack chain 
and generate a security report:

PDDL ATTACK CHAIN:
[insert PDDL sequence here]

REPORT REQUIREMENTS:
1. ATTACK WALKTHROUGH: 
Translate each PDDL action into natural
language, explaining what the attacker 
does at each step.

2. MITIGATION RECOMMENDATIONS: 
Provide specific, actionable defensive
measures. Include concrete commands or 
configuration changes.

Include metadata: 
number of steps, attacker skill level 
required, and overall risk assessment.
\end{verbatim}
\endgroup

\subsection{Output: Generated Natural Language Report}

\textbf{Attack Walkthrough}:
\begin{enumerate}
    \item \textbf{Step 1}: The attacker spawns a Docker process leveraging their group membership.
    
    \item \textbf{Step 2}: Using Docker's host filesystem mounting capability, the attacker modifies the systemd timer file to include a malicious payload while maintaining the file's root execution context.
    
    \item \textbf{Step 3}: When the systemd timer triggers, it executes the modified script with root privileges, granting the attacker a root shell.
\end{enumerate}

\textbf{Mitigation Recommendations}:
\begin{itemize}
    \item \textbf{File Permissions}: Secure systemd timer file:
\begingroup\small
\begin{verbatim}
sudo chown root:root /etc/systemd/system/
        cloud-logger.timer
sudo chmod 644 /etc/systemd/system/
        cloud-logger.timer
\end{verbatim}
\endgroup
    \item \textbf{Group Policy}: Restrict docker group to trusted users only
    \item \textbf{Monitoring}: Implement auditing for systemd file modifications:
\begingroup\small
\begin{verbatim}
sudo auditctl -w /etc/systemd/system/
-p wa -k systemd_files
\end{verbatim}
\endgroup
    \item \textbf{Docker Hardening}: Implement Docker daemon security options:
\begingroup\small
\begin{verbatim}
# In /etc/docker/daemon.json
{
  "authorization-plugins": ["cayley"],
  "userns-remap": "default",
  "no-new-privileges": true
}
\end{verbatim}
\endgroup
\end{itemize}

\textbf{Chain Metadata}:
\begin{itemize}
    \item Steps: 3
    \item Attacker skill: Intermediate (requires Docker knowledge)
    \item Risk level: Critical (direct path from default user to root)
    \item CVSS Estimate: 8.8 
    \item Impact: Complete system compromise
\end{itemize}

\end{document}